\documentclass[aps, prx, 10pt, twocolumn,superscriptaddress]{revtex4-2}

\usepackage{graphicx}
\usepackage{color}
\usepackage{amsmath}

\newcommand{\lifen}{Li$_2$(Li$_{1-x}$Fe$_{x}$)N}

\begin{document}
\title{Cooperative quantum tunneling of the magnetization in Fe-doped Li$_3$N}

\author{M. Fix}
 \affiliation{EP VI, Center for Electronic Correlations and Magnetism, Institute of Physics, University of Augsburg, D-86159 Augsburg, Germany}
\author{J. H. Atkinson}
 \affiliation{Department of Physics, University of Central Florida, Orlando FL 32816, USA}
\author{F. Müller}
 \affiliation{EP VI, Center for Electronic Correlations and Magnetism, Institute of Physics,
 University of Augsburg, D-86159 Augsburg, Germany}
\author{E. del Barco}
 \affiliation{Department of Physics, University of Central Florida, Orlando FL 32816, USA}
\author{A. Jesche}
 \email[]{anton.jesche@physik.uni-augsburg.de}
 \affiliation{EP VI, Center for Electronic Correlations and Magnetism, Institute of Physics, University of Augsburg, D-86159 Augsburg, Germany}

\begin{abstract}
The spin-reversal in dilute \lifen~with $x < 1$\,\% is dominated by resonant quantum tunneling of spatially separated spins.
We report on the effect of finite couplings between those states that give rise to cooperative, simultaneous quantum tunneling of two spins.
This phenomenon, known as spin-spin cross relaxation, effectively elucidates the fine-structure observed in isothermal magnetization loops, a previously unresolved aspect. 
Temperature and field-dependent magnetization measurements were conducted over a range from $T = 0.23$\,K to 300\,K in applied fields of up to $\mu_0H = 7.5$\,T.
The effect of transverse fields is investigated and magnetic dipole fields are computed numerically. 
Our findings affirm the importance of magnetic interactions in dilute \lifen~and underscore its exemplary suitability as a model system for investigating spin-reversal processes at the microscopic level. 
This is attributed to its comparatively simple crystal structure, the availability of large single crystals, elevated characteristic energies, and well-defined energy levels.
\end{abstract}

\maketitle

\section{Introduction}
Controlling spin states in a solid at the atomic level is at the forefront of modern data storage and quantum computing. 
One of the most fascinating ways a spin can change its orientation is by quantum tunneling that takes place not only on microscopic but also on mesoscopic scales\,\cite{Wernsdorfer2002c,Barbara2014}. 
This tunneling process allows for transitions between spin states that are separated by a large barrier although there is insufficient energy for a classical passage. 
Initially discovered in single molecule magnets (SMMs)\,\cite{Thomas1996, Barbara1995, Friedman1996, Hernandez1996}, this effect has since been intensively studied in the context of quantum computation\,\cite{Leuenberger2001}, quantum coherence\,\cite{Bertaina2007}, 
and quantum entanglement\,\cite{Candini2010}. 
Whereas most SMMs are built from clusters of magnetic ions, there are also systems based on the magnetic moment of isolated ions\,\cite{Ishikawa2003, Alam2006, AlDamen2008, Zadrozny2013a} that are referred to as single ion magnets (SIMs). 
Such mono-nuclear magnetic centers were also studied in a comparatively small number of purely inorganic materials\,\cite{Zykin2020}, for example Ho-doped LiYF$_4$\,\cite{Giraud2001},  Cu-doped alkaline-earth phosphate apatites\,\cite{Kazin2014}, or Fe-doped Li$_3$N\,\cite{Jesche2014b, Fix2018c}.

The latter shows quantum tunneling of the magnetization at comparatively high temperatures $T > 10$\,K and extremely strong sensitivity to small applied fields\,\cite{Fix2018c}: $H = 30$\,Oe applied parallel to the easy axis almost fully suppresses quantum tunneling and freezes the orientation of the magnetic moment either parallel or anti-parallel to the easy axis. 
On the other hand, $H = 100$\,Oe applied perpendicular to the easy axis leads to strongly enhanced spin flip probability, a clear indication for resonant tunneling\,\cite{Fix2018c}.
The availability of large single crystals, high characteristic temperatures and energy scales, and structural simplicity make Li$_2$(Li$_{1-x}$Fe$_x$)N an ideal model system to improve our understanding of quantum tunneling and related phenomena. 
X-ray spectroscopy results further reinforce this suitability, revealing that Li$_2$(Li$_{1-x}$Fe$_x$)N is clean of defects and disorder with a random distribution of Fe in the Li$_3$N host matrix\,\cite{Huzan2020}. 
This is corroborated by Moessbauer spectroscopy, which revealed that quantum tunneling is still effective at high temperatures of $T \sim 70$\,K\,\cite{Braeuninger2020}. 

\begin{figure}
\includegraphics[width=0.42\textwidth]{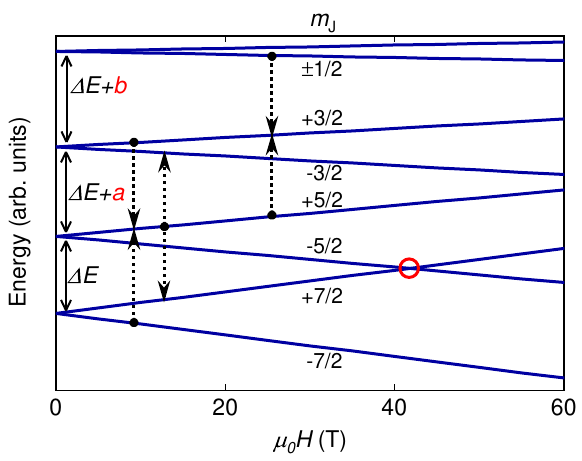}
\caption{Schematic of spin-spin cross relaxation (adapted from\,\cite{Barbara2003}). Depicted are energies of a $J = 7/2$ state that is split into four non-equidistant doublets as a function of an applied magnetic field.  The dotted, black arrows indicate simultaneous transitions of two individual spins without changing the total energy. Quantum tunneling of single spins takes place when (avoided) level crossings of different spin states occur (red circle). For Fe-doped Li$_3$N, this would require huge applied fields of more than $\mu_0 H \approx 42$\,T. \label{sscr_schematic}}	
\end{figure}

However, so far the structure of the isothermal magnetization curves ($M(H)$), which serve as one of the primary sources for analyzing quantum tunneling effects in SMMs and SIMs, was not well understood.
Whereas the large step at $H = 0$ is accurately described by resonant tunneling of isolated Fe moments, there are additional, small but well-defined anomalies, for example at $\mu_0H = 0.13$ and 0.4\,T, that form a complex fine-structure (see below). 
A rough estimate yields a corresponding magnetic energy scale of $0.1-0.5$\,meV, which seems too large for hyperfine couplings but too small for any known electronic interaction \cite{Xu2017}. 
Based on single-atom processes, the first step would require applied fields of several tens of Tesla in order to cause a crossing of energy levels.

Here we show that almost all details of $M(H)$ are well described by cooperative quantum tunneling (of pairs) of spins, which is a manifestation of spin-spin cross relaxation (SSCR)\,\cite{Bloembergen1959, Wernsdorfer2002a}. 
The process is based on a simultaneous transition of two spins under conservation of the total energy (see Fig.\,\ref{sscr_schematic}). 
A weak coupling due to dipolar and/or exchange interactions leads to a collective quantum process\,\cite{Wernsdorfer2002a}. 
We shall discuss the presence of mesoscopic, entangled pairs of spins in \lifen~that extend over several unit cells of the host lattice.

\section{Experimental}\label{2exp}
Single crystals of several millimeters along a side were grown from a lithium-rich flux\,\cite{Jesche2014b,Jesche2014c}. 
Temperature-dependent and isothermal magnetization were measured using a Quantum Design Magnetic Property Measurement System (MPMS3) equipped with a 7\,T magnet. The data obtained were corrected for the diamagnetic sample holder:
The sample was sandwiched between two Torlon discs and fixed inside a straw. The diamagnetic contribution of the Li$_3$N host was subsequently subtracted using $\chi_\mathrm{M}(\mathrm{Li}^{1+}) = -8.8\cdot 10^{-12}$\,m$^3$mol$^{-1}$\,\cite{Banhart1986} and $\chi_\mathrm{M}(\mathrm{N}^{3-}) = -1.63\cdot 10^{-10}$\,m$^3$mol$^{-1}$\,\cite{Hohn2009}.
Magnetization measurements at lower temperatures of $T = 0.23$\,K were performed by a custom-made high-sensitivity micro-Hall effect magnetometer in a He3 Oxford Instruments fridge in the presence of magnetic fields generated by an American Magnetics 3D vector superconducting magnet.
Chemical analysis was performed using inductively coupled plasma optical emission spectroscopy (ICP-OES, Vista-MPX). To this end, the samples were dissolved in a mixture of hydrochloric acid and distilled water.
For simplicity, the nominal Fe concentrations are used through the text. 
The measured values for $x$ are provided in Tab.\,1.

\begin{table}[b]
\caption{\label{tabicp}
Fe concentration $x$ in \lifen~determined by inductively coupled plasma optical emission spectroscopy (ICP-OES).}
\begin{ruledtabular}
\begin{tabular}{ccc}

sample \#	&   $x$ nominal & $x$ from ICP-OES\\
  \hline
1 			& 0.001			& 0.00130(10) 	\\
2 			& 0.003 			& 0.00291(18) 	\\
3 			& 0.005 			& 0.00533(32)	\\
4			& 0.009			& 0.00886(53)	\\
5			& 0.003			& 0.00320(30)	\\
\end{tabular}
\end{ruledtabular}
\end{table}

\section{Results}\label{ch:magnetic_properties}

\begin{figure}
\includegraphics[width=0.47\textwidth]{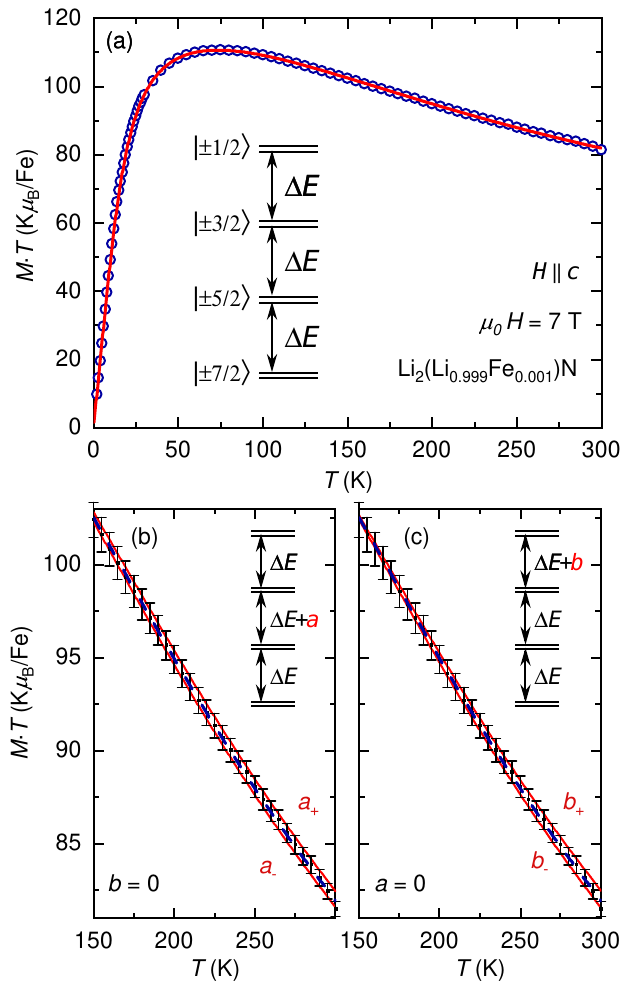}
\caption{Temperature-dependent magnetization of a single crystal of Li$_{2.999}$Fe$_{0.001}$N after field cooling in $\mu_0 H = 7$\,T. 
The solid line in (a) represents a fit according to equation\,\eqref{eq:boltzmann} and yields an energy level separation of 22.4\,meV. The assumed level configuration for $H = 0$ is depicted as inset. 
Panels (b) and (c) show fits to equation\,\eqref{eq:boltzmann} using the parameters obtained from the above fit (blue, dashed line) plus small additional shifts of the energy levels by $a$ and $b$ (red, solid lines) as depicted in the insets. $a_-$ and $b_-$ represent the lower limits of $a$ and $b$ resulting in curves still within the error bar, where $b = 0$ and $a = 0$, respectively. $a_+$ and $b_+$ denote the corresponding upper limits.
\label{M-T}}	
\end{figure}

\subsection{Temperature dependent magnetization}
\label{ch:mvt}
In the following section, we are going to show that the measured magnetic susceptibility is consistent with the presence of non-equidistant doublet states, which is a prerequisite for the occurrence of SSCR.
Figure\,\ref{M-T} shows the temperature-dependent magnetization $M(T)\cdot T$ of a single crystal of Li$_{2.999}$Fe$_{0.001}$N after field cooling in $\mu_0 H = 7$\,T.
The $H$-field was applied parallel to the crystallographic $c$-axis ($H \parallel c$).
$M(T)\cdot T$ is expressed in units of Bohr magneton per Fe (for clarity) and displays a temperature dependence over the whole investigated temperature range ($2{-300}$\,K). 
This is typical for systems with a large magnetic anisotropy and reflects changes in the Boltzmann populations of the relevant energy levels with temperature\,\cite{VanVleck1978, Zadrozny2013c}.

Therefore, the temperature dependence of the magnetization was modeled taking into account explicitly the Boltzmann probabilities of the low-lying magnetic states. 
These can be best described as an effective $J = 7/2$ multiplet, split by spin-orbit coupling into four doublets with quantum numbers $m_J = \pm 7/2, \pm 5/2, \pm 3/2, \pm 1/2$ \cite{Zadrozny2013a, Xu2017} (see Fig.\,\ref{M-T} for a schematic of the level scheme).
Since each state carries the moment $\mu(m_J) = -g m_J \mu_\mathrm{B}$, the resulting total magnetic moment of the system (projected along the applied field) is given by equation\,\eqref{eq:boltzmann}:

\begin{equation}
\begin{gathered}
\label{eq:boltzmann}
\mu (T) = \sum_i \mu_i p_i = \\
=\frac{1}{Z}\sum_{m_J = -J}^J g \mu_\mathrm{B} m_J \exp \left(-\frac{E_{m_J} + g m_J \mu_\mathrm{B} B}{k_\mathrm{B}T}\right),
\end{gathered}
\end{equation}

where $Z$ represents the partition function:

\begin{equation}
Z = \sum_i p_i = \sum_{m_J = -J}^J \exp \left(-\frac{E_{m_J} + g m_J \mu_\mathrm{B} B}{k_\mathrm{B}T}\right),
\end{equation}

and $E_{m_J}$ equals the energy of $|m_J\rangle$ with respect to the ground state doublet $|\pm 7/2\rangle$, $g = 10/7$ is the Landé factor obtained for $J = 7/2$, $L = 2$, and $S = 3/2$\,\cite{Xu2017}.
In a first approach equidistant energy levels, separated by $\Delta E$, 
ware introduced, in accordance with the almost equidistant levels calculated by quantum chemistry methods \cite{Xu2017}. The free parameters in the fit are $\Delta E$, the Fe concentration $x$ and a small, temperature independent offset.
Due to extremely slow relaxation for temperatures below $T=16$\,K\,\cite{Jesche2014b,Fix2018c} the fit was restricted to the interval $[20\,\mathrm{K}, 300\,\mathrm{K}]$.

The solid line in Fig.\ref{M-T} represents a fit of equation\,\eqref{eq:boltzmann} and shows a remarkable agreement with the measured data. 
An energy level separation of $\Delta E = 22.4(2)$\,meV (261\,K) was found.
The Fe content $x=0.12(1)$\,\% obtained from the fit is in excellent agreement with the concentration determined via ICP-OES [$x=0.13(1)$\,\%]. 
The temperature-independent offset converged to a small value close to 0.4\,\% of the low-temperature magnetization. 
Note that the Landé factor of $g = 10/7$ is actually determined for a free ion.
Nevertheless, when releasing $g$ in the fit, it converged to a value that deviates by less than $10^{-4}$ from the Landé factor.

In accordance with small deviations of the calculated levels from equidistance\,\cite{Xu2017}, the parameters $a$ and $b$ (see Fig.\,\ref{sscr_schematic} and schematic inset in Fig.\,\ref{M-T}b,c) were introduced, representing additional energy shifts of the excited states $m_J = \pm 3/2$ and $m_J = \pm 1/2$, respectively. 
The parameters determined from the fit with $a = b = 0$ (see above) were fixed, and $a$ and $b$ were varied separately until the calculated curve showed significant deviations from the measured data (the error is dominated by the weighing error). 
The corresponding curves are shown in Fig.\,\ref{M-T}b,c, where $a_-$, $b_-$ denote the lower and $a_+$, $b_+$ the upper limit of $a$ and $b$, respectively. 
The intervals for $a = [-1.3\,\mathrm{meV}, +1.9\,\mathrm{meV}]$ and $b = [-2.6\,\mathrm{meV}, +3.9\,\mathrm{meV}]$ obtained in this way are slightly asymmetric, indicating a small increase of the level distance with increasing energy, in accordance with the calculation in Ref.\,\cite{Xu2017}. 
As expected, the allowed interval for the shift of the higher excited level $b$ is larger than for the shift $a$ of the lower excited state, reflecting the reduced influence of higher energy levels in the studied temperature range.

\subsection{Isothermal magnetization}

\begin{figure}
    \includegraphics[width=0.48\textwidth]{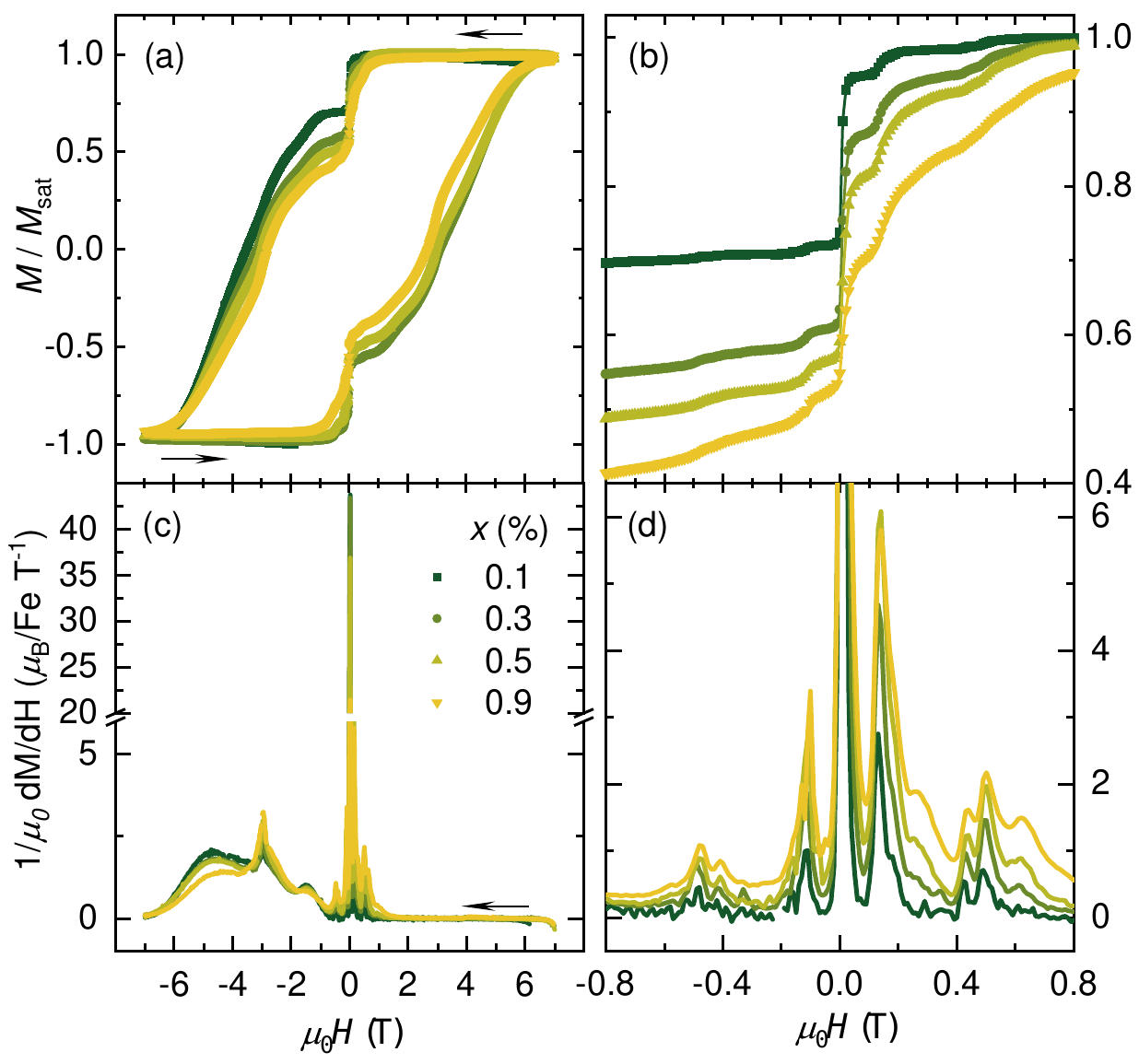}
    \caption{Isothermal magnetization of \lifen~at $T = 2$\,K for samples with varying Fe content $x$. 
Lower panels show the derivative of the magnetization with respect to the applied field.
Panels on the right hand side show an enlarged view around $H = 0$.
As evident from magnetization and derivative, the step-size at $H \approx 0$ decreases with increasing $x$, whereas all other steps increase, indicating a collective nature of the latter ones. 
\label{M-H}}
\end{figure}

Figure\,\ref{M-H} shows isothermal magnetization curves $M(H)$ of \lifen~at $T = 2$\,K and a sweep-rate of $\mu_0\mathrm{d}H/\mathrm{d}t = 0.39$\,mT/s for samples with different Fe content $x$ (samples 1-4).
Arrows indicate the direction of the field sweep.
The $H$-field was applied parallel to the crystallographic $c$-axis. 
In order to obtain saturation, the samples were field cooled prior to the measurement in $\mu_0H = 7$\,T. 
For better comparison, the curves are normalized to their saturation magnetization at $T = 2$\,K [$M_\mathrm{sat} = (5.0\pm 0.3)\,\mu_\mathrm{B}/\mathrm{Fe}$].

The isothermal magnetization curves show pronounced steps at several field values with the largest change in magnetization at $H = 0$ (Fig.\,\ref{M-H}a,b) 
Whereas the step at $H = 0$ shows a decrease in intensity for higher Fe concentrations, all steps in finite fields increase with~$x$.
Even without invoking the microscopic origin of this behavior, it indicates that the step at $H = 0$ is caused by isolated, non-interaction Fe-atoms whereas Fe-Fe interactions are at play at the smaller steps in finite fields.
This behavior is also clearly seen in the derivatives shown in Fig.\,\ref{M-H}c,d. 
Note that we refer to these anomalies also as peaks in the following, which refers to the corresponding derivative d$M$/d$H$.

\begin{figure}
	\includegraphics[width=0.48\textwidth]{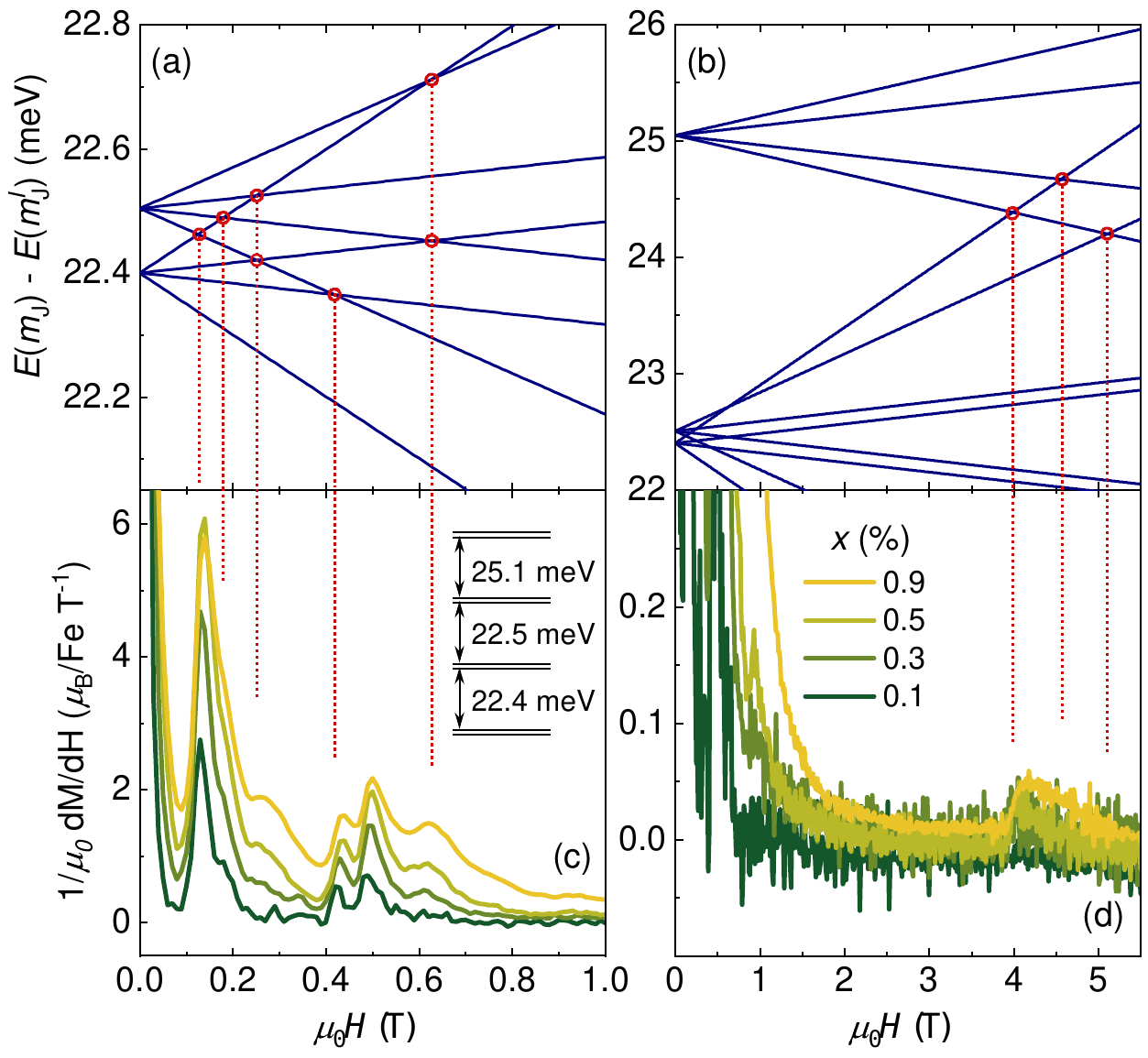}
 \caption{(a) Energy differences between ground state and first exited state (lower branch emerging from 22.4\,meV) and between first and second excited state (upper branch emerging at 22.5\,meV) as well as (b) between second and third excited state as a function of an applied magnetic field.
Possible SSCR transitions are highlighted by red circles.
(c,d) Derivative of $M(H)$ with respect to $H$ showing maxima at crossings of energy level differences ($T = 2$\,K, $H \parallel c$, Fe concentration $x$).  
 \label{deltaE}}
\end{figure}

As shown in Fig.\,\ref{deltaE}, the position of the steps in $M(H)$ can be well explained by the occurrence of identical energy differences between the magnetic states. 
Instead of energy levels as depicted in Fig.\,\ref{sscr_schematic}, the differences between those levels are plotted in Fig.\,\ref{deltaE}a as function of $H$.
Accordingly, the value of 22.4\,meV at $H = 0$ corresponds to the energy splitting between ground state and first excited state, whereas 22.5\,meV is the energy difference between first and second exited state, that is $\Delta E + a$ with $a = 0.1$\,meV. 
Note that this is well within the range of $a = [-1.3\,\mathrm{meV}, +1.9\,\mathrm{meV}]$ obtained from the analysis of $M(T)$, which was presented in the previous section. 
As a function of $H$, several crossings of energy level differences are observed in the field range up to $\mu_0H = 1$\,T. 
Those fit well with the peak positions found in the derivative of $M(H)$, which is plotted in Fig.\,\ref{deltaE}c.
The first crossing at $\mu_0H = 0.13$\,T corresponds to simultaneous transitions $\vert -7/2 \rangle \rightarrow \vert +5/2 \rangle$ and $\vert -3/2 \rangle \rightarrow \vert +5/2 \rangle$.

Another set of anomalies in $M(H)$ is observed around $\mu_0H \approx 4$\,T. 
Those are well described by similar crossings of energy differences assuming $b = 2.7$\,meV that corresponds to an energy gap of $\Delta E = 22.4\,\mathrm{meV} + b = 25.1$\,meV between the second and the third excited state (Fig.\,\ref{deltaE}b). 
Note that this is also well in the allowed range of $b = [-2.6\,\mathrm{meV}, +3.9\,\mathrm{meV}]$ (see previous section). 
The crossing of energy level differences does correspond well with anomalies in $M(H)$ as shown in Fig.\,\ref{deltaE}d.  

The level schemes depicted in Fig.\ref{deltaE}a,b have been calculated based on $g = 10/7$, level splittings of $\Delta = 22.4$\,meV + a/b, and the values of $m_J$ provided above (Fig.\,\ref{M-T}), which allow for a precise description of $M(T)$. 
Nevertheless, it must be approached with caution, as the precise crossings are highly dependent on both the size of the magnetic moments and the exact zero-field splittings. 
More likely than not, the $g$-factors of the doublet states differ somewhat from the values given above.
The value of $a = 0.1$\,meV was chosen such that the preponderance of the data is well described.
However, other combinations are also possible including those that describe the peak in the derivative at $\mu_0H = 0.5$\,T. 
We refrain from presenting these in order to keep the number of correlated parameters low and because of further difficulties that are discussed below.

\subsection{Minor loops}

One of the anomalies in $M(H)$ behaves distinctly different from all the others; the step in the third quadrant at $\mu_0H = -2.96$\,T (see Fig.\,\ref{M-H}a and the corresponding peak in d$M$/d$H$ in Fig.\,\ref{M-H}c).
This peak shows no corresponding anomaly in the first quadrant. 
Furthermore, the peak position does not change with the Fe concentration. 
In fact, we find the peak centered at $\mu_0H = -2.955 \pm 0.007$\,T (error estimated by two times standard deviation) for all samples \#1-4 whereas significant and systematic shifts are observed for all other anomalies (see Fig.\,\ref{peakpositions} and corresponding text in the discussion section). 

\begin{figure}
	\includegraphics[width=0.46\textwidth]{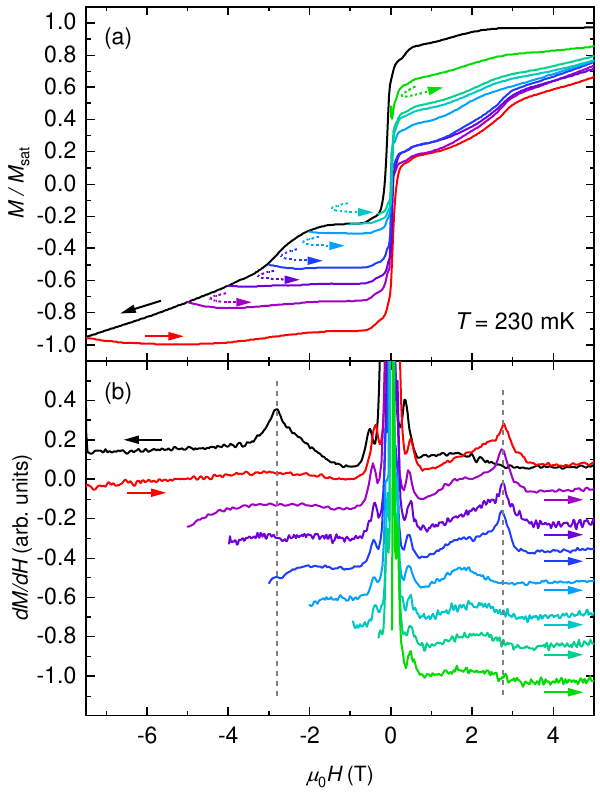}
 \caption{(a) $M(H)$ loops with the reversal of the applied field successively shifted to smaller vales ($T = 230$\,mK, $H \parallel c$, $x = 0.3$\,\%, sample \#5). The black curve presents the initial $M(H)$ for all curves whereas the colored lines show $M(H)$ after reversing the applied field.  
At $\mu_0H = 2.8$\,T a pronounced step-like anomaly is found that vanishes in the first quadrant when the field reversal takes place before crossing the anomaly in the third quadrant. 
(b) Derivative of the $M(H)$ curves shown in (a).
 \label{minorloops}}
\end{figure}

This motivated the measurement of a series of incomplete $M(H)$ (minor loops) where the applied field was reversed prior to reaching saturation. 
The measurements were performed on sample \#5 with $x = 0.0032$ at $T = 0.23$\,K with $H \parallel c$ and a sweep-rate of $\mu_0$d$H$/d$T = 4.2$\,mT/s. 
The obtained results are shown in Fig.\,\ref{minorloops} with $M(H)$ shown in panel a and the corresponding derivatives plotted in panel b.
The black and the red curve present a regular $M(H)$ loop similar to the ones presented in Fig.\,\ref{M-H}a though with a roughly 10 times higher sweep rate.
A clear, step-like anomaly appears in $M(H)$ and a corresponding peak is seen in the derivative d$M$/d$H$ at $\mu_0H = -2.82$\,T (black curves in Fig.\,\ref{minorloops}a,b).
The slight change of the peak position is likely a direct consequence of the higher sweep rate and in line with the position of the central peak (see discussion).
When the field is reversed, the anomaly emerges at $\mu_0H = +2.79$\,T (red curves in Fig.\,\ref{minorloops}a,b).

In a series of further measurements the applied field was reversed at successively smaller values, whereas the initial demagnetization is always given by the black curve (for example the violet curve where the applied field was reversed at $\mu_0H = -5$\,T). 
There is no significant difference to the full loop for the first three minor loops where the field was reversed at $\mu_0H = -5, -4$, and -3\,T, respectively. 
A drastic change, however, is observed when the field is reversed before the anomaly emerges; the light blue curve with a field reversal at $\mu_0H = -2$\,T shows no anomaly in the first quadrant. 
Further runs with early field reversal show identical behavior (see Fig.\,\ref{minorloops}b, three bottom most curves). 

We associate the observed behavior with another magnetic species within the sample, independent of the one causing the peaks at lower fields (which are the focus of the main discussion of this paper). 
This second species shows a negligible tunneling probability at zero field compared to the main species. 
We speculate that this species originates from a small fraction of either strongly coupled neighboring Fe spins that behave as spin dimers, or single Fe ions with a different local site symmetry that prevents the mixing of states necessary to create tunnel splitting leading to magnetization relaxation at zero field.
Furthermore, the presence of a second magnetic species is consistent with high-field electron spin resonance measurements on Co-doped Li$_3$N\,\cite{Albert2021}.

\subsection{Transverse fields}

\begin{figure}
	\includegraphics[width=0.48\textwidth]{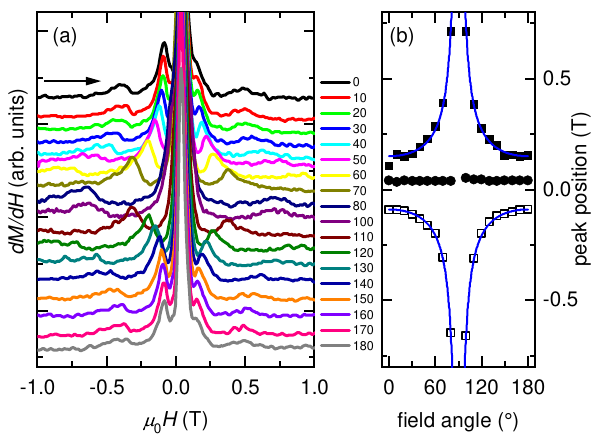}
 \caption{(a) Derivatives of $M(H)$ loops for field applied at an angle with respect to the crystallographic $c$-axis. 
 (b) position of the three main peak shown in a). Open squares: last minor peak before passing $H = 0$, closed squares: first minor peak after passing $H = 0$, closed circles: major peak at $H \approx 0$. Blue lines show fits to the projection of the transverse field (see text).
 \label{transverse}}
\end{figure}

In a previous publication it was shown that the central anomaly close to $H = 0$ is strongly affected by transverse fields whereas the additional, smaller anomalies are not\,\cite{Fix2018c}. 
However, only effects of small applied fields up to $\mu_0H = 40$\,mT had been investigated. 
In this section, we present $M(H)$ loops with large transverse components by applying the field at various angles $\phi$ with respect to the crystallographic $c$-axis (sample \#5, $x = 0.3$\,\%, $T = 4.2$\,K, sweep-rate $\mu_0$d$H$/d$T = 4.2$\,mT/s). 
Results are shown in Fig.\,\ref{transverse} in form of derivatives d$M$/d$H$: for $\phi = 0$ and 180$^\circ$ the minor peaks are in accordance with data shown in Fig.\,\ref{M-H}. For the discussion of the $\phi$-dependence, we focus on the minor peaks closest to the central peak. 
A clear shift with increasing $\phi$ from 0 to 80$^\circ$ is observed that is symmetric with respect to the $ab$-plane. 
It turns out that the peak position is solely determined by the projection of the applied field toward the $c$-axis. 

In order to accurately determine the peak positions, all curves plotted in Fig.\,\ref{transverse}a were consistently fitted by five pseudo-Voigt profiles.    
A plot of the obtained peak positions together with a fit to the equation $H_0/cos(\phi)-H_{\rm off}$ is shown in Fig\,\ref{transverse}b with $H_0$ being the ideal peak position for $\phi = 0$. 
We find $\mu_0H_0 = \pm 0.12(1)$\,T and $\mu_0H_{\rm off} = - 0.033(3)$\,T. 
The latter accounts for a constant offset of the applied field and is in good agreement with the position of the main peak at $\mu_0H = 0.039(4)$\,T. 
In other words, the main peak is centered between the closest minor peaks and the full $M(H)$ loops seem to be shifted by $H_{\rm off}$ independent from $\phi$. 
Furthermore, the analysis indicates that the area of the (first minor) peaks does not dependent on $\phi$, even though a large relative error of 50\% has been estimated (mainly caused by the empirical choice of the fit profile). 
Accordingly, the step-size of the minor anomalies in $M(H)$ seem largely independent from the transverse field, which is in stark contrast to the behavior of the central peak\,\cite{Fix2018c} and in line with the proposed fundamental different origin of the anomalies.

\section{Discussion}\label{discussion}

\begin{figure}
\includegraphics[width=0.42\textwidth]{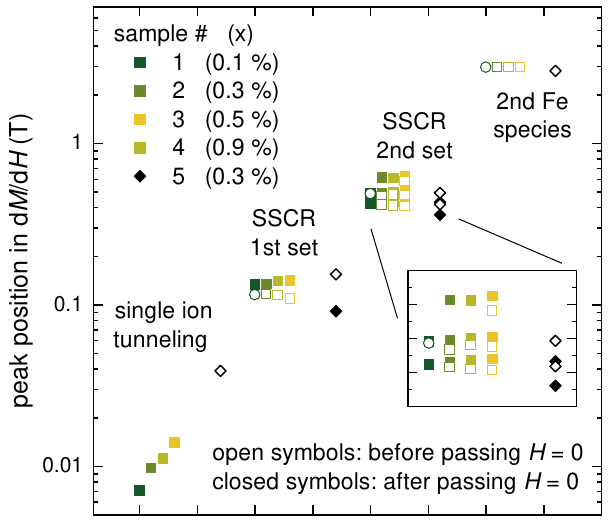}
\caption{Anomalies observed in $M(H)$ loops described by peaks in the derivatives d$M$/d$H$ for samples \# 1-5. The open symbols refer to transitions that appear before the applied field crosses $H = 0$, whereas the closed symbols refer to anomalies that are observed afterwards. 
The arrangement of the points along the horizontal axis only serves to improve the comparability of the different Fe concentrations (without implying a functional dependence).
\label{peakpositions}}	
\end{figure}

We first want to focus on the symmetry of the steps in $M(H)$ with respect to the zero point of the applied field: all anomalies observed in isothermal magnetization measurements are summarized in Fig.\,\ref{peakpositions} for the five samples investigated.
Anomalies that appear before the applied field crosses $H = 0$ are denoted by open symbols whereas anomalies that appear afterwards (and therefor also after the main transition at $H \approx 0$) are shown by closed symbols.  
Accordingly, the latter denote the absolute value of peak positions that are observed at negative applied fields when the initial saturation of $M(H)$ was realized in positive fields.

The major peak (labeled 'single ion tunneling' in Fig.\,\ref{peakpositions}) is not exactly located at $H = 0$ but somewhat shifted with a clear trend for samples \#1-4, which were all measured in the same setup with identical sweep rates of $\mu_0\mathrm{d}H/\mathrm{d}t = 0.39$\,mT/s: the transition shifts to higher fields for increasing the Fe concentration $x$. 
This indicates that a change in the internal field (which is proportional to $x$) causes the main transition to occur at different applied fields. 
Sample \#5 behaves markedly different with the transition occurring before the applied field crosses $H = 0$ even though the $x$ is similar to sample \#2.  
This is likely caused by the different measurement routines: samples \#1-4 were measured in discrete steps with an average sweep rate whereas sample \#5 was measured in 'sweep mode' (using a different experimental setup employing a Hall sensor). 
The observed behavior is not surprising given that the relaxation rates and time dependencies of the magnetization in general are extremely field dependent for $H \approx 0$\,\cite{Fix2018c}.

The anomalies appearing next to the main transition are proposed to be caused by SSCR and labeled 'SSCR 1st set' in Fig.\,\ref{peakpositions}.
As for the main peak, there is a clear dependence on $x$ with an increase of the field difference between those first minor peaks for increasing $x$. 
In fact, the major peak is located close to the center between the first minor peaks, which indicates a shift of the whole $M(H)$ curves with respect to (the zero point of) the applied field (in accordance with section\,\ref{ch:magnetic_properties}D).

The second set of anomalies that appear next to the first minor peaks are labeled 'SSCR 2nd set' in Fig.\,\ref{peakpositions}.
A further analysis of this set is complicated by the strong overlap of several peaks and the lower signal-to-noise ratio. 
However, the same trend as found for the 1st set of anomalies is observed: the distance between two corresponding anomalies follows the shift of the main transition (see inset).

Denoted by 'coupled flip' in Fig.\,\ref{peakpositions} are the anomalies discussed in section\,\ref{ch:magnetic_properties}C. There is no significant dependence on $x$ for samples \#1-4 (see section\,\ref{ch:magnetic_properties}B), however, sample \#5 shows the transition at a somewhat smaller field of $\mu_0H = -2.82$\,T, which is likely caused by the different field sweep mode. 

In all previous publications, quantum tunneling in \lifen~had been described in the framework of non-interacting spins.
However, a finite coupling (by dipolar or exchange interactions) between the spins is necessary for SSCR to emerge and suitable Hamiltonians have been set up and diagonalized for SMMs (see for example \cite{Wernsdorfer2002a}).
Applying those established models to \lifen~is complicated by the following issues:  
\\a) significant mixing of the eigenstates of the free ion is expected. 
Even in the simplest picture, the crystal electric field mixes $\vert \pm 7/2 \rangle$ and $\vert \pm 5/2 \rangle$ states in the hexagonal symmetry of the Fe atom\,\cite{Segal1970}. 
\\b) There is significant mixing between Fe $3d$ and $4s$ states\,\cite{Novak2002, Huzan2024}
\\c) In contrast to most of the SMMs, the zero-field splitting of \lifen~cannot be described by a uniaxial anisotropy constant that contributes with $-D S_z^2$ to the Hamiltonian. Instead, a more complex energy level scheme is present\,\cite{Xu2017}. 
\\d) The simultaneous presence of a large crystal electric field and an unquenched orbital moment\,\cite{Jesche2014b}.
Those characteristic properties of \lifen~make it - in the authors opinion - impossible to take advantage of established schemes to find and diagonalize an effective Hamiltonian.   

For these reasons, the focus of this paper will be on the experimental results.
Nevertheless, a preliminary analysis suggests that the major difficulty of SSCR in such a single-ion picture, that is a negligible occupation of energy levels more than 22\,meV above the ground state at $T = 2$\,K, is resolved in a more rigorous treatment, which could include the possible presence of phonon superradiance\,\cite{chudnovsky_superradiance_2002, chudnovsky_phonon_2004}.

SSCR effects similar to the ones presented here were observed in SMMs, for example in complexes containing clusters of Mn\,
\cite{Wernsdorfer2002a,Wernsdorfer2005,Wernsdorfer2004b,Milios2006}, 
Ni\,\cite{YangEn2006,Hameury2013}, 
or Fe ions\,\cite{Vergnani2012,Compain2009,Cornia2019}. 
Furthermore, the effect was demonstrated in single crystals of LiYF$_4$ doped with Ho ions\,\cite{Giraud2001,Giraud2003b,Barbara2004}.
Typically in those reports, the vast majority of steps in $M(H)$ is attributed to resonant tunneling of isolated spins and SSCR transitions present smaller, additional anomalies.
In stark contrast, there is only one single-ion transition in \lifen~but several associated with SSCR. 
The reason is the significantly larger anisotropy energy of the title compound that causes the (avoided) level crossing to appear only at large applied fields of several tens of Tesla. 
An experimental verification seems challenging since the relaxation is slow compared to the pulses at high-magnetic-field facilities\,\cite{Fix2018b,Fix2018c}. 
Accordingly, the SSCR transitions in \lifen~are more pronounced since, besides the zero-field transition, they provide the only spin-reversal mechanism for $\mu_0H < 10$\,T.

Similar to our findings, the step size of SSCR transitions in $M(H)$ was found to increase with increasing concentration of magnetic ions as shown for the SMM [(Pc)$_2$Ho$_{0.02}$Y$_{0.98}$]$^-$TBA$^+$ when increasing the Ho concentration\,\cite{Ishikawa2005}.
It remains to be seen, whether the transverse field dependence of SSCR in \lifen~is smooth when compared to single atom transitions\,\cite{Wernsdorfer2002a}.

Finally, we are going to elaborate on the strength of the coupling between the Fe magnetic moments.
Since a combinatorial analysis\,\cite{Klatyk2002} is cumbersome for such low Fe concentrations, a numerical approach was chosen by placing Fe atoms on the lattice of Li$_3$N according to various Fe concentrations $x$ using the NumPy library\,\cite{Harris2020} for Python.
A $40 \times 40 \times 40$ lattice has been simulated over 1000 times. 
Since SSCR is a two-particle process between pairs of Fe atoms, we focus on the field created by the nearest neighbor (n.n).
Magnetic dipole fields $B_{\rm dip}$ were calculated for a fully polarized state with all magnetic moments ($\mu = 5\,\mu_B$) pointing along the crystallographic c-axis.
The results are summarized in Table\,\ref{tabdip}.
The first value represents the largest possible dipole field that is caused by an Fe atom placed on a nearest neighbor Fe site along the $c$-axis. 
Basically, even larger values are possible for chains of Fe atoms.
However, those are extremely unlikely to form for $x < 0.01$.
For Fe atoms sitting on the in-plane n.n. site, the largest possible, negative dipole field amounts to -99\,mT.
Due to the six in-plane n.n., the probability of being occupied is somewhat larger: for $x = 0.9$\,\% it amounts to 5.3\,\%, whereas it drastically reduces to 0.6\,\% for the smallest Fe concentration of $x = 0.1$\,\%.

The average field caused by the nearest neighbor for $x = 0.1$\,\% is $B_{\rm dip} = 2.1$\,mT along the $c$-axis and $B_{\rm dip} = 1.7$\,mT perpendicular to the $c$-axis. 
The standard deviation is significantly larger than the average since $B_{\rm dip}(r)$ is highly non-linear and the average is dominated by a small number of particularly large values.
Nevertheless, those results reveal that a significant number of Fe atoms are subject to magnetic dipolar fields in the range of 10\,mT.
Larger values for the average $B_{\rm dip}$  were obtained for $x = 0.9$\,\% that roughly scale with the Fe concentration. 

Furthermore, the simulation allows to draw conclusions on the single-atom transition at $H \approx 0$.
To this extent, we calculated the ratio of Fe atoms that are subject to $B_{\rm dip} < 3$\,mT, which was found to be the threshold for quantum tunneling to appear\,\cite{Fix2018c}.
The ratio decreases from 92(5)\,\% for $x = 0.1$\,\% to 56(4)\,\% for $x = 0.9$\,\%.
This scales well with a decrease of the step size in $M(H)$ by roughly 1/2 (Fig.\,\ref{M-H}c).

\begin{table}
\caption{\label{tabdip}
Magnetic dipole fields at the position of Fe atoms caused by nearest neighbor (n.n.) Fe magnetic moments in a fully polarized state. For the smallest and the largest Fe concentration $x$, the averaged, absolute values are provided with the standard deviation given in brackets.}
\begin{ruledtabular}
\begin{tabular}{ccc}

  &   $B_{\rm dip} \parallel c$ (mT)			& $B_{\rm dip} \perp c$ (mT)\\
  \hline
 n.n. along 	$c$				& 169 			& 0 			\\
 n.n. in-plane 				&  -99  		 	& 0 			\\
 $\emptyset$ n.n. $x = 0.1$\,\%	& 2.1 (6.7) 		& 1.7 (4.1)	\\
 $\emptyset$ n.n. $x = 0.9$\,\%	& 12 (27)		& 7.3 (12)	\\
\end{tabular}
\end{ruledtabular}
\end{table}

\section{Summary}
This study investigates quantum tunneling of the magnetization in \lifen, a model system due to the comparatively simple crystal structure, the availability of large single crystals, sharp energy levels and high characteristic energy scales (with respect to anisotropy energy, relaxation rates, and crossover to the quantum tunneling regime). 
The temperature-dependent magnetization of dilute \lifen~can be satisfactorily described as a result of the magnetic moment of single, isolated Fe ions. 
Through detailed measurements of isothermal magnetization, the research uncovers complex spin transitions that deviate from conventional non-interacting spins. 
Instead, the driving force behind the observed magnetic anomalies is identified as cooperative quantum tunneling of spin pairs, known as spin-spin cross relaxation (SSCR).
In particular, it is shown that the observed behavior is not caused by structural defects. 
We believe that this work represents a further important step in understanding the complex magnetic behavior of the structurally rather simple compound \lifen~and will improve our understanding of spin-reversal processes on a microscopic scale.

\section*{Acknowledgments}
We thank Alexander Herrnberger and Klaus Wiedenmann for technical support, Andrea Moos for performing ICP-OES.
Helpful comments provided by Thilo Kopp are gratefully acknowledged. 
This work was supported by the Deutsche Forschungsgemeinschaft (DFG, German Research Foundation) - JE 748/1.


%

\end{document}